\documentclass[12pt]{article}
\usepackage{a4wide,epsfig,psfrag,amsmath,amssymb,cite,scalefnt}
\usepackage{color}

\graphicspath{ {figs/} }

\newcommand{\num}[1]{\color{black} #1}

\allowdisplaybreaks

\begin{document}

\title{\vskip-3cm{\baselineskip14pt
    \begin{flushleft}
     \normalsize TTP16-055, MITP/16-144
    \end{flushleft}} \vskip1.5cm 
  Four-loop photon quark form factor and cusp anomalous dimension
  in the large-$N_c$ limit of QCD
  }

\author{
Johannes Henn$^{a}$,
Roman N. Lee$^{b,e}$,
Alexander V. Smirnov$^{c}$,
\\
Vladimir A. Smirnov$^{d,e}$,
Matthias Steinhauser$^{e}$,
\\[1em]
{\small\it (a) PRISMA Cluster of Excellence, Johannes Gutenberg University,}\\
{\small\it  55099 Mainz, Germany}
\\
{\small\it (b) Budker Institute of Nuclear Physics,}\\
{\small\it 630090 Novosibirsk, Russia}
\\
{\small\it (c) Research Computing Center, Moscow State University}\\
{\small\it 119991, Moscow, Russia}
\\  
{\small\it (d) Skobeltsyn Institute of Nuclear Physics of Moscow State University}\\
{\small\it 119991, Moscow, Russia}
\\
{\small\it (e) Institut f{\"u}r Theoretische Teilchenphysik,
 Karlsruhe Institute of Technology (KIT)}\\
 {\small\it 76128 Karlsruhe, Germany}  
}
  
\date{}

\maketitle

\thispagestyle{empty}

\begin{abstract}

  We compute the four-loop QCD corrections to the massless
  quark-anti-quark-photon form factor $F_q$ in the large-$N_c$ limit.  From
  the pole part we extract analytic expressions for the corresponding cusp
  and collinear anomalous dimensions. 

  \medskip

  \noindent
  PACS numbers: 11.15.Bt, 12.38.Bx, 12.38.Cy

\end{abstract}

\thispagestyle{empty}


\newpage


\section{Introduction}

Perturbation theory is a powerful tool to obtain reliable predictions for
physical observables within the Standard Model of particle physics or its
extensions. Due to the high precision of experimental results, e.g., at the
CERN Large Hadron Collider (LHC) or at the $B$ factories, it is on the one
hand mandatory to advance the development of tools, which can be used for
higher order calculations. On the other hand it is necessary to improve the
understanding of the perturbative structure of quantum field theories.  Form
factors are ideal objects to obtain deeper insight into the latter.
Especially in the context of QCD they are indispensable tools to investigate
the infrared structure of scattering amplitudes to high orders in perturbation
theory. Moreover, from the pole part it is possible to extract universal
process-independent
quantities~\cite{Mueller:1979ih,Collins:1980ih,Sen:1981sd,Magnea:1990zb,Korchemsky:1991zp,Korchemskaya:1992je,Sterman:2002qn}
like the cusp anomalous dimension which can be extracted from the
$1/\epsilon^2$ pole of the form factor.
The finite parts of the form factors serve as building blocks for a variety of
physical processes. For example, the quark-anti-quark-photon form factor
enters the virtual corrections of the Drell-Yan process for the production
of lepton pairs at hadron colliders.

In this paper we consider the quark-anti-quark-photon form
factor which is conveniently obtained from the photon-quark vertex function
$\Gamma^{\mu}_q$ by applying an appropriate projector. In $D=4-2\epsilon$
space-time dimensions we have
\begin{eqnarray}
  F_q(q^2) &=& -\frac{1}{4(1-\epsilon)q^2}
  \mbox{Tr}\left( p_2\!\!\!\!\!/\,\,\, \Gamma^\mu_q p_1\!\!\!\!\!/\,\,\,
  \gamma_\mu\right)
  \,,
\end{eqnarray}
with $q=p_1+p_2$ where $p_1$ and $p_2$ are the incoming quark and anti-quark
momenta and $q$ is the momentum of the photon.  We perform our calculation in
the framework of QCD keeping the number of colours, $N_c$, generic. In the
limit of large $N_c$ the calculation of $F_q$ is simplified since only planar
Feynman diagrams contribute. This is the limit we consider in this paper.
Besides $N_c$ we also keep the number of active quark flavours, $n_f$ as
a parameter and thus have at four-loop order the colour structures
$N_c^4$, $N_c^3 n_f$, $N_c^2 n_f^2$, $N_c n_f^3$ where each factor of $n_f$
counts the number of closed fermion loops.

Two- and three-loop corrections to $F_q$ have been computed in
Refs.~\cite{Kramer:1986sg,Matsuura:1987wt,Matsuura:1988sm,Gehrmann:2005pd,Baikov:2009bg,Gehrmann:2010ue,Lee:2010ik,Gehrmann:2010tu}.
To obtain the four-loop corrections two obstacles need to be overcome: (i) the
reduction to a set of basis integrals and (ii) the (if possible) analytic
calculation of the latter.  Recently, the first steps towards four loops have
been initiated by computing the fermionic contributions to $F_q$ in the planar
limit~\cite{Henn:2016men}. The $n_f^3$ terms have been confirmed in
Ref.~\cite{vonManteuffel:2016xki} using different methods both for the
reduction and the computation of the master integrals.  Let us mention that
the four-loop corrections to the cusp anomalous dimension with two and three
closed fermion loops have also been obtained in Ref.~\cite{Davies:2016jie}.
In Ref.~\cite{Boels:2015yna} one finds a discussion of non-planar master
intergrals relevant for the four-loop form factor in a ${\cal N}=4$
supersymmetric Yang-Mills theory.

In the present paper, we evaluate the $n_f^0$
contribution and therefore complete the evaluation of the form factors and
anomalous dimensions in the limit of large $N_c$.

In the next Section we provide some technical details, in particular to
the calculation of the most complicated master integral, and we discuss our
results in Section~\ref{sec::res}. We provide explicit expressions
for the four-loop cusp and collinear anomalous dimensions in the planar limit.
Furthermore, we provide results for the finite part of $\log(F_q)$.


\section{Technical details}

\begin{figure}[t] 
  \begin{center}
    \includegraphics[width=0.3\textwidth]{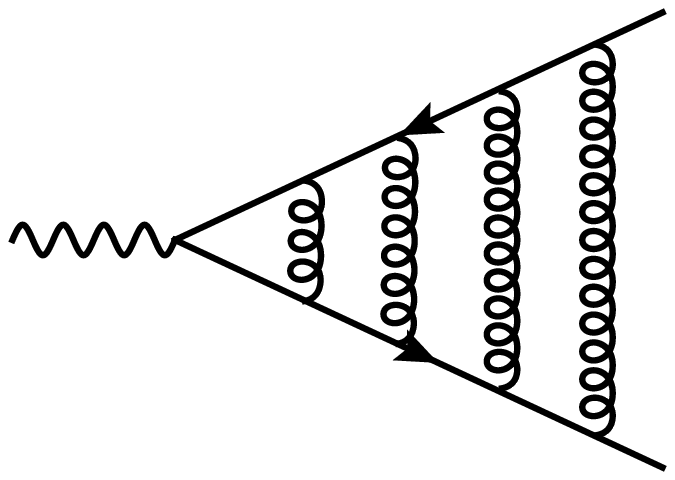}  \hfill
    \includegraphics[width=0.3\textwidth]{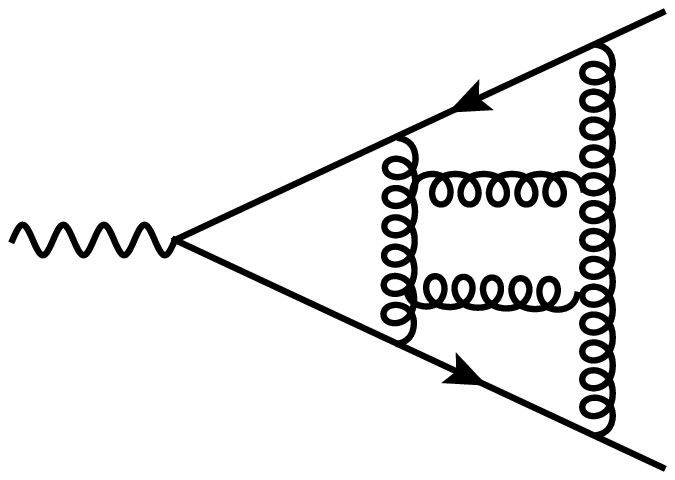}  \hfill
    \includegraphics[width=0.3\textwidth]{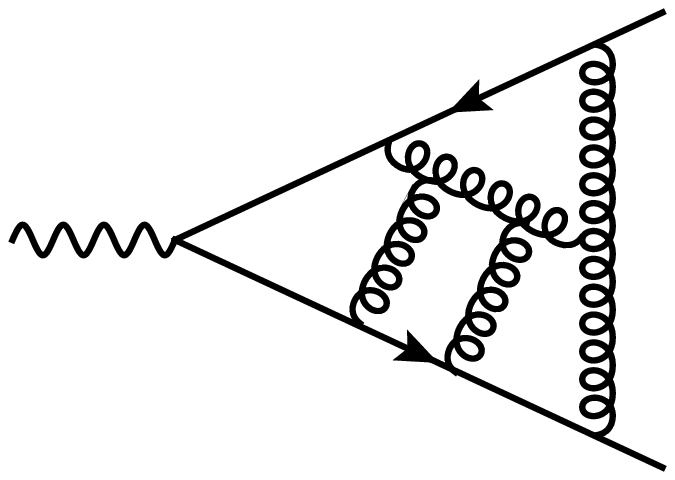}
    \caption{\label{fig::diags}Sample four-loop Feynman diagrams contributing to $F_q$.
      Solid, curly and wavy lines represent
      quarks, gluons and photons, respectively. All particles are massless.}
  \end{center}
\end{figure}

For the calculation of $F_q$ we use a well-tested automated chain of programs
which work hand-in-hand. The Feynman amplitudes are generated with {\tt
  qgraf}~\cite{Nogueira:1991ex}. Since there is no possibility to select
already at this point the planar diagrams also non-planar amplitudes are
generated and we obtain in total 1, 15, 337 and 9784 diagrams at one, two,
three and four loops.  Sample Feynman diagrams at four loops can be found in
Fig.~\ref{fig::diags}.  Next, we transform the output to {\tt
  FORM}~\cite{Kuipers:2012rf} notation using {\tt q2e} and
{\tt exp}~\cite{Harlander:1997zb,Seidensticker:1999bb}.  The
program {\tt exp} furthermore maps each Feynman diagram to predefined integral
families for massless four-loop vertices with two different non-vanishing external
momenta; 68 of them are of planar type.  At this point we perform the Dirac
algebra and decompose the numerator into terms which appear in the
denominator. This allows us to express each Feynman integral as a linear
combination of scalar functions which belong to the corresponding family.
After exploiting the symmetries connected to the exchange of the external
momenta we can reduce the number of families, for which integral tables have to
be generated, from 68 to 38. Note that for the fermionic contributions, which
have been considered in Ref.~\cite{Henn:2016men}, only 24 families are needed.

For the reduction to master integrals we use the 
program {\tt FIRE}~\cite{Smirnov:2008iw,Smirnov:2013dia,Smirnov:2014hma} which we apply
in combination with {\tt LiteRed}~\cite{Lee:2012cn,Lee:2013mka}. We observe
that the non-fermionic diagrams lead to more complex integrals for which
the reduction time significantly increases. 
Let us remark that we have adopted Feynman gauge for the calculation
of the non-fermionic parts.

\begin{figure}[t] 
  \begin{center}
    \includegraphics[width=0.5\textwidth]{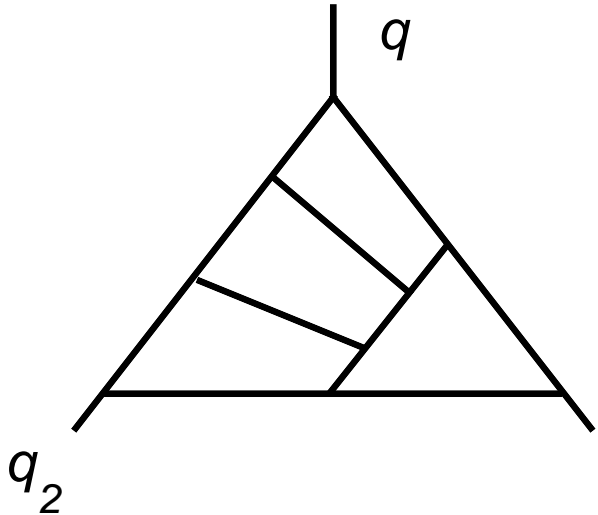}
    \caption{\label{fig::I99}Massless four-loop vertex diagram with two
      external (incoming) momenta $q$ and $q_2$ with $q^2\not=0\not=q_2^2$ and
      $(q+q_2)^2=0$. The master integral $I_{99}$ corresponds
      to $q_2^2=0$.}
  \end{center}
\end{figure}

Once reduction tables for each family are available we apply {\tt tsort},
which is part of the latest {\tt FIRE} version~\cite{Smirnov:2014hma}. It is
based on ideas presented in Ref.~\cite{Smirnov:2013dia} to establish relations
between between primary master integrals and thus minimize their number. In
this way we arrive at {\num 99} master integrals.  
In the remainder of this section we describe in detail the calculation of the
most complicated integral corresponding to the graph of Fig.~\ref{fig::I99}, $I_{99}$.

To have the possibility to apply differential equations for the evaluation of
the form factor master integral corresponding to the graph of
Fig.~\ref{fig::I99} we introduce a second mass scale by considering $q_2^2 = x
q^2$ and derive differential equations with respect to the ratio of the two
scales, $x$. This strategy was advocated in \cite{Henn:2013nsa} and used for
the four-loop form-factor integrals in~\cite{Henn:2016men,HSS}. We derive
differential equations for the corresponding 332 master integrals using {\tt
  LiteRed}~\cite{Lee:2012cn,Lee:2013mka}.

To solve our differential equations we use
an important observation made in Ref.~\cite{Henn:2013pwa}. 
It has been suggested to turn from the basis of primary master
integrals to a so-called canonical basis where the corresponding integrals
satisfy a system differential equations which has a particular structure: the
dependence on $\epsilon$ appears as a linear prefactor and the matrix in front
of the vector of master integrals has only simple poles in $x$, i.e. has only
so-called Fuchsian singularities. Such a system can then easily be solved  
in terms of iterated integrals.

In Ref.~\cite{Henn:2013pwa}, it was proposed that choosing integrals with
  constant leading singularities provides a canonical basis of the
  differential equations. In subsequent work, this was used in a variety of
  cases, see
  e.g. Refs.~\cite{Henn:2014qga,Gehrmann:2014bfa,Mastrolia:2014wca}.  As was
  motivated in Ref.~\cite{Henn:2016men}, this choice can be done in a
  systematic way at the level of the loop integrand.  The first algorithm to
  convert a given differential system to a canonical form at the level of the
  differential equations has been provided in the case of one
  variable\footnote{The very recent paper~\cite{Meyer:2016slj} makes some
    interesting progress towards the algorithmic reduction in the multivariate
    case.} in Ref.~\cite{Lee:2014ioa} (therein called $\epsilon$-form, see
  also Ref.~\cite{Henn:2014qga}).
Recently a public implementation~\cite{Gituliar:2016vfa}
of the algorithm of Ref.~\cite{Lee:2014ioa} in a computer code {\tt Fuchsia}
became available. In this paper we follow~\cite{Lee:2014ioa} to construct a
canonical basis. An independent evaluation of $I_{99}$ can be found in
  Ref.~\cite{HSS} where the canonical basis was chosen at the level of the loop
  integrand.  The size of the system is large but, as it concerns the
diagonal blocks, their size is at most $5\times5$. Consequently, the reduction
of diagonal blocks within the approach of Ref.~\cite{Lee:2014ioa} is
simple. Note that already after this step one might claim that the solution is
expressible in terms of harmonic polylogarithms (and also construct this
solution). Nevertheless, we follow the prescription of Section~7 of
Ref.~\cite{Lee:2014ioa} to reduce the whole system to an
$\epsilon$-form. Similar to Ref.~\cite{Lee:2016lvq} we find that the
differential equation-based hierarchy of the set of the master integrals is
too restrictive and the use of a sector-based hierarchy when factoring
$\epsilon$ out of the whole matrix is necessary.

The family of one-scale Feynman integrals (with $q_2^2=0$ and with $q^2\neq
0$) corresponding to Fig.~\ref{fig::I99} contains {\num 76} master
integrals. After introducing $q_2^2\not=0$ we define $x=q_2^2/q^2$ and obtain
a family of Feynman integrals with {\num 332} master integrals.  Our strategy
is to turn from a primary basis to a canonical basis, solve differential
equations for the canonical basis, evaluate the naive values of the elements
of the canonical basis at $x=0$ (i.e., setting $x=0$ under the integral
sign) from which it will be straightforward to obtain analytical results for
the primary master integrals of our one-scale family, in particular, $I_{99}$.

Following Ref.~\cite{Lee:2014ioa} we arrive at a canonical basis $g$ which is obtained
from the primary basis $f$ by a linear transformation with a matrix $T$,
\begin{eqnarray}
  f &=&  T\cdot g\,.
  \label{eq::T}
\end{eqnarray}
The vectors $f$ and $g$ have $332$ entries and $T$ is a $332\times 332$ matrix.
The dependence of $f$, $g$ and $T$ on $x$ and $\epsilon$ has been suppressed.

It is convenient \cite{Henn:2013pwa} to normalize the canonical master integrals such that 
they have uniformly transcendental $\epsilon$-expansion which starts from $\epsilon^0$.
In our four-loop case, one has to compute 
expansion terms including $\epsilon^8$ terms and we have
\begin{eqnarray}
  g(x,\epsilon) &=& \sum_{k=0}^{8} g_k(x) \epsilon^k
  \,.
\end{eqnarray}
The canonical basis $g$ satisfies (by definition) a
differential equation of the form
\begin{eqnarray}
  g^\prime(x,\epsilon) &=& \epsilon \, A(x)\cdot g(x,\epsilon)
  \,,
  \label{eq::dgl}
\end{eqnarray}
where
\begin{eqnarray}
  A(x) &=& \frac{a}{x} + \frac{b}{x-1}
  \,,
\end{eqnarray}
with constant matrices $a$ and $b$.  It is straightforward to construct the
generic solution of~(\ref{eq::dgl}) order-by-order in $\epsilon$ in terms of
harmonic polylogarithms
(HPLs) \cite{Remiddi:1999ew} with letters 0 and 1, with $332\times 9$ unknown constants.

To fix these constants we use boundary conditions at the point $x=1$.  The
primary master integrals are regular at this point where the integrals become
propagator-type integrals which are well known~\cite{Baikov:2010hf}.  In
particular, all the corresponding 28 master integrals are known
analytically~\cite{Baikov:2010hf,Lee:2011jt} in an $\epsilon$-expansion up to
weight 12 and have been cross checked numerically~\cite{Smirnov:2010hd}.  We
obtain the boundary values of the elements of the canonical basis $g$ at $x=1$
by inverting Eq.~(\ref{eq::T}) and considering the limit $x\to1$. The matrix
$T^{-1}$ involves elements which develop poles up to order $1/(1-x)^6$. This
requires that the first seven expansion terms for $x\to1$ of the primary
master integrals $f$ have to be computed. The corresponding reduction tables
are again generated with {\tt FIRE}. Note that the resulting two-point
integrals with external momentum $q$ have scalar products in the numerator
involving the momentum $q_2$ which complicates the calculation.  It is an
important cross check of the calculation that all poles in $1/(1-x)$ cancel in
the combination $T^{-1} \cdot f$ and we obtain the values for $g$ at $x=1$, in
an $\epsilon$-expansion up to $\epsilon^8$. This fixes the solution of the system
of differential equations in Eq.~(\ref{eq::dgl}).

In a next step we analyze the leading asymptotic behaviour of $g$ near $x=0$.
On the one hand, we obtain it from the differential equations~(\ref{eq::dgl})
where the term $b/(x-1)$ on the right-hand side can be neglected and the
solution has the form $g(x,\epsilon) = h(\epsilon) x^{\epsilon a}$.  The
quantity $x^{\epsilon a} = e^{\epsilon \log(x) a}$ can be evaluated using the
{\tt Mathematica} command {\tt MatrixExp[]} which leads to a $332\times 332$
matrix where each element is a linear combination of terms $x^{k\epsilon}$
with integer $k$.  In general both non-positive $k$ and positive $k$ might
appear. However, in the case of Feynman integrals only terms with
non-positive $k$ can be present
which we use as a check.  On the other hand, the leading asymptotic
behaviour in the limit $x\to0$ can also be obtained with the help of the
{\tt Mathematica} package {\tt HPL}~\cite{Maitre:2005uu} from our analytic
expression for the canonical basis.  Matching the two expressions provides
values for the vector $h(\epsilon)$ in an $\epsilon$-expansion up to
$\epsilon^8$ and terms $x^{k\epsilon}$ with $k=0,-1,-2,\ldots$.  Note
that this step involves powers of $\log(x)$ terms; their cancellation in the
matching provides a welcome check for our calculation.  Finally, the naive
value for $g(x,\epsilon)$ at $x=0$ is obtained by setting all terms
$x^{k\epsilon}$ with $k\not=0$ to zero in the expression for $x^{\epsilon a}$.

In the last step, using Eq.~(\ref{eq::T}), we compute the naive values of
the elements of the primary basis $f$ from the naive expansion of 
the canonical basis $g(x)$ near $x\to0$.
Note that some of the matrix elements of $T$ involve singularities up
to order $1/x^3$. Thus, the naive expansion of $g(x)$ up to order $x^3$ is
needed. It is obtained following the prescription outlined in
Ref.~\cite{Wasow} where the expansion terms can be computed from 
the leading order asymptotics at $x\to 0$
after recursively solving matrix equations with $332\times 332$ entries.
After inserting the expansion of $g$ in Eq.~(\ref{eq::T}) the poles
cancel and the naive values of the primary master integrals at $x=0$ are
obtained. The naive value of one of the elements of our primary basis $f$
is nothing but the one-scale master integral $I_{99}$ (cf. Fig.~\ref{fig::I99}) 
for which we obtain the following analytic result 
\begin{eqnarray}
  I_{99} &=&
  e^{4\epsilon\gamma_E} \left(\frac{\mu^2}{-q^2}\right)^{4\epsilon}  \Bigg\{
%
%
\frac{1}{\epsilon^7}  \Bigg[
-\frac{1}{288}
\Bigg]
%
%
+\frac{1}{\epsilon^6}  \Bigg[
\frac{13}{576}
\Bigg]
%
%
+\frac{1}{\epsilon^5}  \Bigg[
-\frac{101}{576}-\frac{\pi ^2}{48}
\Bigg]
\nonumber\\&&\mbox{}
+\frac{1}{\epsilon^4}  \Bigg[
-\frac{17 \zeta _3}{54}+\frac{5 \pi ^2}{36}+\frac{145}{96}
\Bigg]
%
%
+\frac{1}{\epsilon^3}  \Bigg[
\frac{1775 \zeta _3}{432}-\frac{767 \pi ^4}{17280}-\frac{5 \pi ^2}{8}-\frac{1669}{144}
\Bigg]
\nonumber\\&&\mbox{}
+\frac{1}{\epsilon^2}  \Bigg[
-\frac{83}{72} \pi ^2 \zeta _3-\frac{21899 \zeta _3}{864}-\frac{3659 \zeta _5}{360}+\frac{31333 \pi ^4}{103680}+\frac{659 \pi ^2}{288}+\frac{11243}{144}
\Bigg]
\nonumber\\&&\mbox{}
+\frac{1}{\epsilon}  \Bigg[
-\frac{40231 \zeta _3^2}{1296}+\frac{745 \pi ^2 \zeta _3}{288}+\frac{18751
  \zeta _3}{144}+\frac{50191 \zeta _5}{360}-\frac{277703 \pi
  ^6}{2177280}-\frac{14015 \pi ^4}{10368}
\nonumber\\&&\mbox{}
-\frac{149 \pi ^2}{24}-\frac{22757}{48}
\Bigg]
\nonumber\\&&\mbox{}
+ \Bigg[
\frac{39173 \zeta _3^2}{324}-\frac{77399 \pi ^4 \zeta _3}{25920}+\frac{4013
  \pi ^2 \zeta _3}{432}-\frac{259559 \zeta _3}{432}-\frac{568 \pi ^2 \zeta
  _5}{45}-\frac{1123223 \zeta _5}{1440}
\nonumber\\&&\mbox{}
-\frac{2778103 \zeta _7}{4032}+\frac{3129533 \pi ^6}{4354560}+\frac{28201 \pi ^4}{5760}+\frac{173 \pi ^2}{36}+\frac{382375}{144}
\Bigg]
\nonumber\\&&\mbox{}
+\epsilon \Bigg[
\frac{4931 s_{8a}}{30}+\frac{2615}{144} \pi ^2 \zeta _3^2-\frac{276671
  \zeta _3^2}{2592}-\frac{2702413 \zeta _5 \zeta _3}{1080}+\frac{154037 \pi ^4
  \zeta _3}{31104} 
\nonumber\\&&\mbox{}
-\frac{55327 \pi ^2 \zeta _3}{432}+\frac{1100461 \zeta _3}{432}+\frac{205 \pi
  ^2 \zeta _5}{9}+\frac{155029 \zeta _5}{48}+\frac{2732549 \zeta
  _7}{1008}-\frac{665217829 \pi ^8}{1306368000}
\nonumber\\&&\mbox{}
-\frac{131003 \pi ^6}{45360}-\frac{747929 \pi ^4}{51840}+\frac{2995 \pi ^2}{36}-\frac{2005247}{144}
\Bigg]
\Bigg\}\,,
\end{eqnarray}
where $\zeta_n$ is Riemann's zeta function evaluated at $n$ and
\begin{eqnarray} 
  s_{8a} &=& \zeta_8 + \zeta_{5,3} \approx 1.0417850291827918834\,.
\end{eqnarray}
$\zeta_{m_{1},\dots,m_{k}}$ are multiple zeta values given by
\begin{eqnarray}
  \zeta_{m_{1},\dots,m_{k}} &=&
  \sum\limits _{i_{1}=1}^{\infty}\sum\limits
  _{i_{2}=1}^{i_{1}-1}\dots\sum\limits _{i_{k}=1}^{i_{k-1}-1}\prod\limits
  _{j=1}^{k}\frac{\mbox{sgn}(m_{j})^{i_{j}}}{i_{j}^{|m_{j}|}}
  \,. 
\end{eqnarray}

As by-product we also obtain analytic results for the remaining {\num 75}
one-scale master integrals and we find agreement with the results obtained in
Ref.~\cite{HSS}.  This constitutes a further cross check for our procedure.
We want to stress that the calculation which is outlined in this section is
largely independent from the one performed in Ref.~\cite{HSS}.


\section{\label{sec::res}Results}

This section is devoted to the analytic results of 
the cusp and collinear anomalous dimensions and the finite
part of $F_q$. Generic formulae where the pole part of $F_q$ is parametrized
in terms of the cusp and collinear anomalous dimensions and the QCD beta function
can, e.g., be found in Refs.~\cite{Becher:2009qa,Gehrmann:2010ue}. In what
follows we use Eq.~(2.3) of Ref.~\cite{Henn:2016men} which displays the pole
parts of $\log(F_q)$ up to four-loop order. In this formula it is assumed
that the one-loop coefficient of the beta function is given by
\begin{eqnarray}
  \beta_0 &=& \frac{11 N_c}{3}-\frac{2 n_f}{3}
  ,
\end{eqnarray}
with $n_f$ being the number of active quarks and
the coefficients of the anomalous dimensions
are defined through
\begin{eqnarray}
  \gamma_x &=& \sum_{n\ge0} \left(\frac{\alpha_s{(\mu^2)}}{4\pi}\right)^n
  \gamma_x^n
  \,,
  \label{eq::gamma_x}
\end{eqnarray}
with $x\in\{\text{cusp},q\}$ and $\alpha_s$ is the renormalized coupling
constant with $n_f$ active flavours.  From Eq.~(2.3) of
Ref.~\cite{Henn:2016men} one observes that the four-loop corrections of
$\gamma_{\rm cusp}$ follows from the $1/\epsilon^2$ term of $\log(F_q)$ and
$\gamma_q$ from the linear pole terms.

In the following we start with explicit results for the
cusp and collinear anomalous dimensions. The four-loop corrections
to $\gamma_{\rm cusp}$ reads
\begin{eqnarray}
  \gamma_{\rm cusp}^3 &=&
  \left(-\frac{32 \pi ^4}{135}+\frac{1280 \zeta _3}{27}-\frac{304 \pi
      ^2}{243}+\frac{2119}{81}\right) N_c n_f^2+\left(\frac{128 \pi ^2
      \zeta    _3}{9} +224 \zeta _5 -\frac{44 \pi^4}{27} 
  \right.\nonumber\\&&\left.\mbox{}
    -\frac{16252 \zeta _3}{27} +\frac{13346 \pi ^2}{243}-\frac{39883}{81}\right) N_c^2    n_f+\left(\frac{64 \zeta
      _3}{27}-\frac{32}{81}\right)    n_f^3
+\left(-32 \zeta _3^2
  \right.\nonumber\\&&\left.\mbox{}
-\frac{176 \pi ^2 \zeta _3}{9}+\frac{20992 \zeta
    _3}{27}-352 \zeta _5-\frac{292 \pi ^6}{315}+\frac{902 \pi
    ^4}{45}-\frac{44416 \pi ^2}{243}+\frac{84278}{81}\right)
    N_c^3
  \,.
\nonumber\\
\label{eq::gamma_cusp3}
\end{eqnarray}
We note that, after taking into account the difference between
  fundamental and adjoint representation of the external fields, 
  the leading transcendental piece of this expression agrees with the result
  for planar ${\cal N}=4$ super Yang-Mills~\cite{Bern:2006ew,Beisert:2006ez}, in agreement with
  expectations from~\cite{Kotikov:2004er}.
Note that $\gamma_{\rm cusp}^3$ entering the $1/\epsilon^2$ pole of $\log(F_q)$
is multiplied by $C_F$. For this reason there is only a $N_c^3$ factor in
front of the $n_f$-independent term in Eq.~(\ref{eq::gamma_cusp3}).  The
one-, two- and three-loop corrections in the large-$N_c$ limit can be found in
Eq.~(2.6) of Ref.~\cite{Henn:2016men} where also the fermionic part of
$\gamma_{\rm cusp}^3$ is shown.  The four-loop coefficient of the collinear
anomalous dimension is given by
\begin{eqnarray}
  \gamma_q^3 &=&
   N_c^3 \left[\left(-\frac{680 \zeta _3^2}{9} -\frac{1567 \pi
    ^6}{20412} +\frac{83 \pi ^2 \zeta
    _3}{9}  +\frac{557 \zeta _5}{9} +\frac{3557 \pi ^4}{19440} -\frac{94807 \zeta _3}{972} +\frac{354343 \pi
    ^2}{17496}
\right.\right.\nonumber\\&&\left.\left.\mbox{}
+\frac{145651}{1728}\right)
    n_f\right]+\left(-\frac{8
    \pi ^4}{1215} -\frac{356 \zeta _3}{243}-\frac{2 \pi ^2}{81}+\frac{18691}{13122}\right) N_c
    n_f^3+\left(-\frac{2}{3} \pi ^2 \zeta _3
\right.\nonumber\\&&\left.\mbox{}
+\frac{166 \zeta _5}{9}+\frac{331 \pi ^4}{2430} -\frac{2131 \zeta
    _3}{243} -\frac{68201 \pi
    ^2}{17496}-\frac{82181}{69984}\right) N_c^2 n_f^2
\nonumber\\&&\mbox{}
+ N_c^4
   \left(\frac{1175 \zeta _3^2}{9}+\frac{82 \pi ^4 \zeta _3}{45}-\frac{377 \pi
    ^2 \zeta _3}{6}+\frac{867397 \zeta _3}{972}+24 \pi ^2 \zeta _5-1489 \zeta
    _5+705 \zeta _7
\right.\nonumber\\&&\left.\mbox{}
+\frac{114967 \pi ^6}{204120}-\frac{59509 \pi
    ^4}{9720}-\frac{120659 \pi ^2}{17496}-\frac{187905439}{839808}\right)
  \,.
\label{eq::gamma_q}
\end{eqnarray}
The one-, two- and three-loop corrections and the fermionic four-loop terms to
$\gamma_q^3$ are listed in Eq.~(2.7) of Ref.~\cite{Henn:2016men}.
The $N_c^4$ term is new.

Finally, we also present the finite part of the form factor. We parametrize
the perturbative expansion in terms of the renormalized coupling constant
and set $\mu^2=-q^2$. Furthermore, it is
convenient to consider $\log(F_q)$ which leads to more compact
expressions. Thus, we have the following parametrization
\begin{eqnarray}
  \log(F_q) &=& 
  \sum_{n\ge1}
  \left(\frac{\alpha_s}{4\pi}\right)^n
  \log(F_q)|^{(n)}
  \,.
\end{eqnarray}
In the  large-$N_c$ limit the four-loop term reads
\begin{eqnarray}
\lefteqn{  \log(F_q)|^{(4)}_{\mbox{\tiny large-$N_c$, finite part}} = } \nonumber\\
&&
   N_c^4 \left(-14 s_{8a}+10 \pi ^2 \zeta _3^2-\frac{86647 \zeta
    _3^2}{54}+766 \zeta _5 \zeta _3-\frac{251 \pi ^4 \zeta
    _3}{6480}-\frac{57271 \pi ^2 \zeta _3}{1296}+\frac{173732459 \zeta
    _3}{23328}
\right.\nonumber\\&&\left.\mbox{}
    +\frac{1517 \pi ^2 \zeta _5}{216}-\frac{881867 \zeta
    _5}{1080}-\frac{36605 \zeta _7}{288}+\frac{674057 \pi
    ^8}{5443200}-\frac{135851 \pi ^6}{77760}+\frac{386729 \pi
    ^4}{31104}
\right.\nonumber\\&&\left.\mbox{}
-\frac{429317557 \pi
    ^2}{839808}-\frac{54900768805}{6718464}\right)
+ \ldots
\end{eqnarray}
where the ellipses refer to the fermionic contributions which are given
in Eq.~(2.8) of Ref.~\cite{Henn:2016men}.
For convenience of the reader we provide the results for the form factor
$F_q$ expanded in the bare strong coupling constant in an ancillary file
which can be downloaded
from~\verb|https://www.ttp.kit.edu/preprints/2016/ttp16-055/|.
This file also contains the lower-loop results expanded to higher order in
$\epsilon$. Furthermore, it contains the dependence of the renormalization
scale $\mu$.


\section{\label{sec::concl}Conclusions and outlook}

We compute the photon-quark form factor to four loops up to the finite term in
$\epsilon$ in the large-$N_c$ limit which is obtained from the planar Feynman
diagrams. From the pole parts we extract the cusp and collinear anomalous
dimensions. We discuss in detail the calculation of the most complicated
master integral (see Fig.~\ref{fig::I99}) and present analytic results
expanded in $\epsilon$ up to transcendental weight eight. An independent
  calculation of this integral, together with a discussion of the
  remaining master integrals can be found in Refs.~\cite{HSS,Henn:2016men} to the
required order in $\epsilon$. We want to remark that the same master integrals
enter the Higgs gluon form factor in the planar limit. However, the
corresponding reduction is significantly more complicated.  The logical
next step is the calculation of the non-planar contributions. 
Note that also here both the reduction and the computation of the
master integrals turn out to be much more complicated.



\section*{\label{sec::ack}Acknowledgments}
We are grateful to Gang Yang for checking numerically the highest poles of our
result for $I_{99}$.  J.M.H. is supported in part by a GFK fellowship and
  by the PRISMA cluster of excellence at Mainz university. This work is
supported by the Deutsche Forschungsgemeinschaft through the project
``Infrared and threshold effects in QCD''.


{{\bf Note added}}:\\
After submission of our mansucript to {\tt arxiv.org} we were
contacted by colleagues drawing our attention to a talk by
Ben Ruijl at the University of Zurich where the result for
$\gamma_{\rm cusp}^3$ has been shown. The result agrees with our Eq.~(11).

\end{document}